\documentclass[pra,twocolumn,amssymb]{revtex4}
\usepackage{graphicx}

\begin{document}
\title{Vortex Dipole in a Trapped Atomic Bose-Einstein Condensate}
\author{ Qi Zhou, Hui Zhai  \\ \it Center for Advanced Study, Tsinghua University, Beijing, China}
\date{\small\today}
\begin{abstract}
We calculate the angular momentum and energy of a vortex dipole in
a trapped atomic Bose-Einstein condensate. Fully analytic
expressions are obtained. We apply the results to understand a
novel phenomenon in the MIT group experiment\cite{Ketterle1}, an
excellent agreement is achieved, and further experimental
investigation is proposed to confirm this vortex dipole mechanism.
We then suggest an effective generation and detection of vortex
dipole for experimental realization. Application of the sum rule
to calculate collective mode frequency splitting is also
discussed.

PACS number:
\end{abstract}

\maketitle

\section{Introduction}
The distinguished behavior of superfluid to carry angular momentum
have attracted people's attention from long ago. It is well known
that vortex excitation plays an important role when a superfluid
responds to the external rotation. Nowadays, the vigorous
development in cold atom physics has provided a new and more
favorable arena to prove many theoretical predictions of rotating
superfluid's behavior. It has been clearly demonstrated by many
recent experiments that the vortex will be a stable excitation
with quite long life-time when sufficient angular momentum is
imparted, and the vortex lattice will be generated when even more
angular momentum is brought to this system.

Besides the vortex itself, the vortex dipole, i.e., a bound
vortex-antivortex pair, is also of particular interest. It is well
known that the low-lying behavior of two-dimensional helium
superfluid is dominated by vortex dipole, for the excitation
energy of a vortex dipole is much smaller than a single vortex.
With the increase of temperature, the vortex pairs will become
unbounded and vortices will be free. This will lead to a phase
transition known as Kosterlize-Thouless(KT) transition. Although
no vortex-antivortex pair has yet been observed directly so far in
atomic Bose-Einstein condensate(BEC), it is natural to presume
that vortex dipole may play an important role when the imparted
angular momentum is not large enough to support the vortex
excitation. So it is worthwhile to study the properties of vortex
dipole excitation in an atomic BEC.

The existence and importance of vortex dipole in an atomic BEC are
also indicated by some recent experiments and numerical
simulations. MIT group stirred the BEC locally with the stirring
frequency which was too small to generate any surface mode,
however, they observed a large number of vortices after stirring
for some time. This suggested a local mechanism of generating
vortex, say through vortex dipole excitation\cite{Ketterle1}.
Recently some numerical calculation also indicated that stable
vortex dipole could exist in a trapped atom BEC for a broad range
of parameters\cite{stablevortexdipole}. All of these lead us to
believe that cold atom physics will offer people opportunity to
investigate this intricate vortex dipole excitation in superfluid.
What is more, the characteristics of trapped atomic BEC will add
much richer physics content to the properties of vortex dipoles,
as well as to the experimental detection and production of vortex
dipoles, which will be emphasized in this paper.

To proceed in this interesting and meaningful research field, we
of course should first make clear the properties of a single
vortex dipole in BEC, i.e., its angular momentum and energy. In
section II, we will first deduce an exact result of its angular
momentum and energy. It is not trivial to calculate the angular
momentum and energy of vortex dipole in an inhomogeneous
condensate. Here we use an integration method in complex
coordinate to obtain fully analytical results, which will be
described explicitly in this section and the appendix. We then
apply our results to investigate MIT group
experiment\cite{Ketterle1}. Through our calculations, we provide
strong evidence to support the presence of vortex dipole during
this kind of vortex nucleation process, and suggest further
experiment to confirm this mechanism.

For experimental study the properties of vortex dipole in details,
an effective way to generate and detect it is required. Although
various methods have been widely adopted to produce vortices
successfully, the formation of vortex dipole has not been
systematically discussed. In section III, we will suggest that
JILA's method\cite{Cornell}, which generate vortex through a
coherent spatial and temporal coupling between two-component BEC,
could also be used to produce vortex dipole.

Measuring the density distribution could demonstrate the density
depression in the neighborhood of the vortex center but have
nothing to do with its phase. The same problem will be encountered
in the case of vortex dipole. We could not distinguish which
vortex is positive or negative only from the density image. What
is more, the velocity field of a vortex dipole decreases quickly
from the dipole center, and any closed path, which is a little far
from the vortex dipole and includes both the vortex and
antivortex, will be quite similar with the case that there is no
singularity at all in this area. It is also not so effective to
use interference method to reveal the phase information of vortex
dipole as in vortex case\cite{Ketterle3}. However,
experimentalists now could accurately  measure the frequency
splitting of the collective modes when the time reversal symmetry
is broken due to the presence of vortices. This splitting will
clearly distinguish vortex dipole from single vortex and
vortex-free states. In theory, there are usually two ways to
calculate the frequency of the collective mode in presence of
vortices, one is perturbative calculation and the other is the sum
rule approach. The results obtained from these two approaches are
in good agreement with each other when the vortex is in the center
of condensate. However, they are no more consistent when the
vortex is off-center as pointed out in Ref.\cite{offaxisvortex}.
In section IV, we first use perturbation approximation to
calculate the frequency splitting of the collective mode, and then
we explain why the sum rule approach fails here. A brief
conclusion and discussion will be given in the last section.

\section{Properties of A Vortex Dipole}

\subsection{Angular Momentum and Energy}

\begin{figure}[htbp]
\begin{center}
\includegraphics[width=1.2in]
{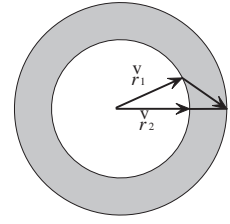} \caption{Vortex dipole configuration: the grey
torus expresses the effective part of the condensate which carries
the angular momentum}\label{fig:dipoleconfig}
\end{center}
\end{figure}

We hereafter consider a BEC confined in a pancake harmonic trap,
which allows us to deal with a quasi-two-dimensional system. The
Gross-Pitaevski equation in a non-rotating coordinate frame is
\begin{equation}
-\frac{\hbar^2}{2m}\nabla^2\psi+V\psi+g\left|\psi\right|^2\psi=E\psi,\label{G-P}
\end{equation}
where $V=\frac{1}{2}m\omega_\bot^2r^2$, and $g$ is the effective
quasi-two-dimensional interaction strength, obtained by
integrating the usual three dimensional one
${4\pi\hbar^2a_{sc}}/{m}$ over the $z$ direction. As we are
considering the small angular momentum case, we could start with
the Thomas-Fermi(T-F) approximation. So the density distribution
of the ground state is well known as
\begin{equation}
\rho(r)=\frac{\hbar\omega_\bot^2{m}}{2g}(R^2-r^2),\label{TFdensity}
\end{equation}
where $R$ is the T-F radius. Once a vortex dipole appears, in the
region outside of vortex core we can write
\begin{equation}
\psi=\sqrt{\rho(r)}e^{i\theta(r,\varphi)}, \label{wavefunction}
\end{equation}
with
\begin{equation}
\theta(r,\varphi)=\Theta(\vec{r}-\vec{r}_1)-\Theta(\vec{r}-\vec{r}_2).\label{dipolephase}
\end{equation}
Here $\Theta(\vec{r})$ is the angle of $\vec{r}$ with respect to
$x$ axis. $\vec{r}_1$ and $\vec{r}_2$ denote the positions of
vortex and antivortex respectively. Eq.(\ref{wavefunction})
follows from the assumption that the presence of vortices will not
change the density distribution of the condensate but only
introduce a phase, except inside the vortex core. This is true
when the vortex core $\Lambda$ is small in the T-F region. With
Eq.(\ref{wavefunction}) we could calculate the angular momentum of
the condensate as
\begin{eqnarray}
L=\hbar\int_0^R\rho{r}dr\int_0^{2\pi}\frac{\theta(r,\varphi)}{\partial\varphi}d\varphi
=\hbar\int_0^R\rho{r}dr\theta(r,\varphi)|^{2\pi}_0.
\end{eqnarray}
Only in the region $r_1<r<r_2$, $\theta$ will change $2\pi$ after
going along a circle. Thus,
\begin{equation}
L=2\pi\hbar\int_{r_1}^{r_2}\rho{r}dr.
\end{equation}
we notice that the energy of a vortex dipole increases with their
distance, and the most efficient way to carry angular momentum is
that $\vec{r}_1$ and $\vec{r}_2$ direct along the same radius,
namely they are fully polarized, as are schematically shown in
fig(\ref{fig:dipoleconfig}). So we could denote
$r_1=D-d/2$,$r_2=D+d/2$, therefore
\begin{equation}
L=2\pi\hbar\rho_0R^2(\frac{Dd}{R^2}-\frac{D^3}{R^3}\frac{d}{R}-\frac{d^3}{4R^3}\frac{D}{R}),\label{L}
\end{equation}
where $\rho_0$ is ${\hbar\omega_\bot^2mR^2}/{(2g)}$.
Now we turn to the energy of the vortex dipole. In the T-F region,
it could be written as
\begin{equation}
E=\frac{\hbar^2}{2m}\int\rho\upsilon^2dxdy,\label{Ed}
\end{equation}
where
\begin{equation}
\vec{\upsilon}=\frac{\hat{z}\times(\vec{r}-\vec{r}_1)}{\left|\vec{r}-\vec{r}_1\right|^2}
-\frac{\hat{z}\times(\vec{r}-\vec{r}_2)}{\left|\vec{r}-\vec{r}_2\right|^2}.
\end{equation}
Here we have neglected the presence of mirror dipole due to the
edge effect, which would be a good approximation in a broad region
except when the dipole is very close to the boundary of the
condensate. For convenience, we do all the calculation in complex
coordinates. We will use $z=x+iy$ and $\bar{z}=x-iy$, and then
obtain
\begin{equation}
E=\frac{\hbar^2\rho_0}{4imR^2}\int_\Omega(R^2-\left|z\right|^2)
\frac{\left|z_1-z_2\right|^2}{\left|z-z_1\right|^2\left|z-z_2\right|^2}d\bar{z}\wedge{dz}.\label{Dipoleenergy}
\end{equation}
According to the Green formula
\begin{equation}
\int_\Omega{d\omega}=\int_{\partial\Omega}\omega,
\end{equation}
we could obtain the energy expression in the region $\Lambda\ll d
\ll R$
\begin{equation}
E=\frac{2\hbar^2\rho_0}{m}\left[1-\left(\frac{D}{R}\right)^2\right]\ln\frac{d}{\Lambda}.\label{E}
\end{equation}
The detailed calculations are given in Appendix B. As the energy
is dependent on $d$ logarithmically, the vortex and antivortex
should bind up to each other closely, and we could neglect higher
orders of $d/R$. Compared with the well known result for
homogenous case, we also find that in Eq.(\ref{E})the homogenous
density is replaced by local density at the position of vortex
dipole. It can be easily understood because the amplitude of the
velocity field of vortex dipole falls rapidly from the center of
the vortex dipole, and most contribution to the energy of a dipole
comes from the local velocity field.

\subsection{Interpretation of Experiment}

It is of long standing interest to investigate the intrinsic
mechanism of vortex nucleation since the study of helium
superfluid. In Ref.\cite{Ketterle1} dynamic nucleation of vortices
in a trapped atomic BEC was studied experimentally. When the
condensate was stirred by a laser with the beam waist comparable
to the T-F radius, the condensate was globally rotated. In this
case they found an enhanced vortex generation when the stirring
frequency coincides that of surface excitations. It is indicated
that vortex could be generated through surface excitation.
However, when the condensate was stirred locally by a small beam,
and the stirring frequency was far below any surface excitation
frequency, they also found large number of vortices. This
phenomenon suggested that the vortex can also be created locally
in the bulk of the condensate, which seems to be in conflict with
the topological argument, unless the vortex dipole was excited as
an intermediate step. With the increase of angular momentum, the
anti-vortices will finally be expelled out from the system and
leave only the vortices in the condensate.

A remarkable feature of the local nucleation is that the vortex
number produced by this local stirring is strongly dependent on
the distance $l$ between stirring position and the center of
condensate. The vortex number presents a maximum in the region
where the ratio of $l$ to the condensate size $R$ is around $0.4$.
Hence, a question can be raised that whether we can understand the
maximum from the properties of vortex dipole excitation.

Our answer to this question is, it is just in this region that the
excitation energy of vortex dipole takes a minimum. In other word,
the vortex dipole is easiest to be excited in this region. With
Eq.(\ref{L}) and Eq.(\ref{E}) in hand, we could calculate the
energy as a function of the position of vortex dipole $D$ for a
given small value of angular momentum $L_0$. The result is
\begin{equation}
\frac{E}{E_{0}}=\left(1-\left(\frac{D}{R}\right)^2\right)\ln\left[
\frac{L_0}{2\pi\hbar\rho_{0}R^2}\frac{R}{\Lambda(D)}\frac{1}{\frac{D}{R}-(\frac{D}{R})^3}\right],\label{ED}
\end{equation}
where $E_{0}$ denoting $2\hbar^2\rho_{0}/m$ is introduced as an
energy unit. Here we should emphasize that the vortex core size
$\Lambda$ also depends on $D/R$ via
\begin{equation}
\Lambda(D)=\frac{\Lambda_{0}}{\sqrt{1-\left(\frac{D}{R}\right)^2}}.
\end{equation}

\begin{figure}[htbp]
\begin{center}
\includegraphics[width=3.0in]
{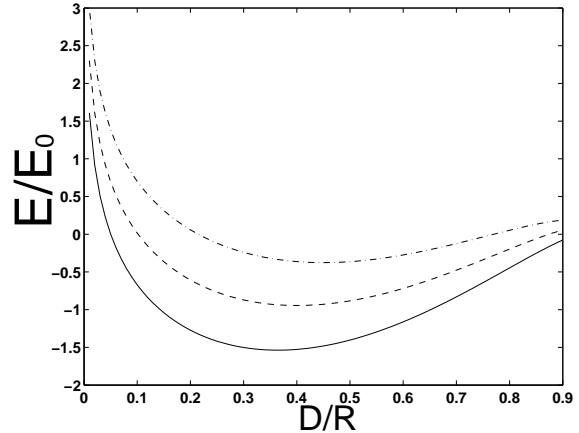} \caption{Vortex dipole energy. Solid, dashed, and dash
dot denote $T=0.05,0.1,0.2$ respectively, where
$T=L_0/2\pi\hbar\rho_0{R}\Lambda_0$}\label{fig:E1}
\end{center}
\end{figure}

This result is plotted in Fig(\ref{fig:E1}). We could clearly see
that the energy minimum occurs at $D/R$ around $0.4$. Our
calculation thus provides a support to the vortex dipole
mechanism.

To further confirm the vortex dipole mechanism, we investigate the
properties of vortex dipole in an anharmonic trap
\begin{equation}
V(r)=\frac{1}{2}m\omega_{\bot}^2r^2+\frac{1}{2}mk^2r^4.
\end{equation}
The angular momentum and energy of the vortex dipole will become
\begin{equation}
L=2\pi\hbar\rho_0R^2\left[(1+A)\frac{Dd}{R^2}-\frac{D^3}{R^3}\frac{d}{R}-A\frac{D^5}{R^5}\frac{d}{R}\right],
\end{equation}
and
\begin{equation}
E=\frac{2\hbar^2\rho_0}{m}\left[1-\left(\frac{D}{R}\right)^2+A\left(1-\left(\frac{D}{R}\right)^4\right)\right]\ln\frac{d}{\Lambda}.
\end{equation}
where $A=\frac{k^2R^2}{\omega_\bot^2}$.
\begin{figure}[htbp]
\begin{center}
\includegraphics[width=3.0in]
{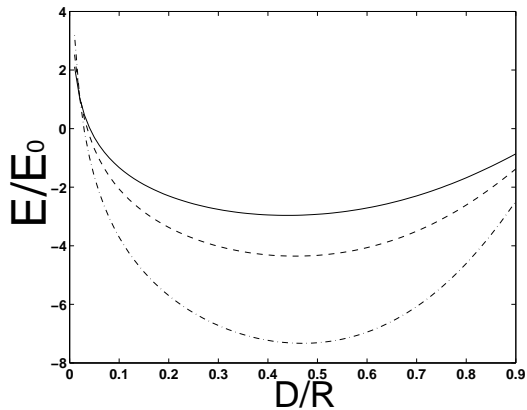} \caption{Vortex dipole energy. Solid, dashed, and dash
dot denote $A=0.5,1,2$ respectively, and $T$ is fixed at
0.05}\label{fig:E2}
\end{center}
\end{figure}
We find that with the increase of the quartic term, the minimum in
the $E-D$ curve moves toward larger value of $D/R$ as shown in
fig.(\ref{fig:E2}. Hence we predict that the stirring position
where the number of generated vortices reaches its maximum will
also change correspondingly, if a similar experiment is performed
in an anharmonic trap.

\section{Vortex dipole formation}
Let us first briefly review the vortex formation by phase
imprinting method\cite{Cornell}, which used a laser to couple a
two-component BEC. When we write the order parameter $\psi$ as
$n^{1/2}e^{i\alpha}\chi$, where
$\chi^T=(\cos(\theta/2)e^{-i\varphi/2},\sin(\theta/2)e^{i\varphi/2})$,
we could map the wave function to a pseudo-spin $\chi$ on a Bloch
sphere. Under the optical coupling, the pseudo-spin will rotate
with the effective Rabi oscillation frequency
$\Omega_{\text{eff}}$. In the experiment the laser was locally
applied and moved around a circle with the frequency $\omega_r$,
and the condensate in different points acquires a phase difference
proportional to $\Omega_{\text{eff}}\Delta{t}$, where $\Delta{t}$
is the time interval of the laser passing through these two
points. If $\omega_r=\Omega_{\text{eff}}$, the phase difference
between the two points is identical to $\Delta\theta$, and the
phase configuration of a vortex is built up. Besides, if we only
change the sign of $\Omega_{\text{eff}}$ and keep the circling
direction of the laser unchanged, an anti-vortex will be produced.

\begin{figure}[htbp]
\begin{center}
\includegraphics[width=3.6in]
{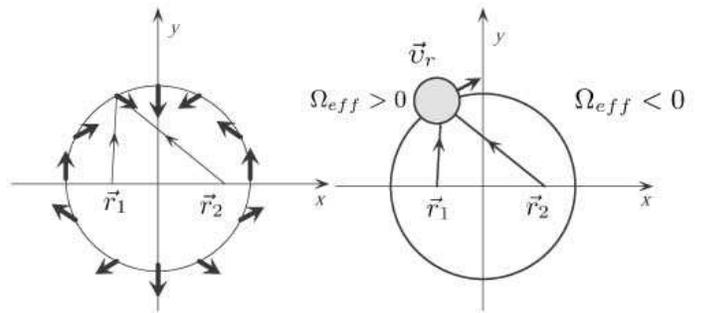} \caption{Vortex dipole generation. Left:
arrowhead denotes the velocity field of the vortex dipole along
the circle. Right:$\vec{\upsilon}_r$ denotes the velocity of the
moving laser}\label{fig:generation}
\end{center}
\end{figure}

Extending the above method to dipole generation is
straightforward. From Fig.(\ref{fig:generation}), we can see that
the velocity field of a vortex dipole is clockwise in left plane
and anti-clockwise in right plane. To produce a vortex dipole, the
moving laser is required to change the sign of
$\Omega_{\text{eff}}$ once it crosses the $y$ axis. What is more,
the phase configuration of the vortex dipole could be produced
exactly. $\Omega_{\text{eff}}$, the velocity of the moving laser
$\vec{\upsilon}_r$ and the phase change gradient of the vortex
dipole $\nabla\theta$ should satisfy the relationship
\begin{equation}
\vec{\upsilon}_r\cdot\vec{\nabla}\theta=\Omega_{\text{eff}}
\end{equation}
everywhere along the circle. $\theta$ denotes dipole phase
expression Eq.(\ref{dipolephase}).
Once the phase along the circle is fixed, according to the
uniqueness of solution to the equation $\nabla^2\theta=0$, the two
phase singularities, as well as the phase elsewhere inside the
circle will consequently be determined.

\section{Vortex dipole detection}

When the ground state of the condensate possesses time reversal
symmetry, the $\pm{m}$ collective modes will be degenerate.
However, once the time reversal symmetry is broken due to the
presence of vortex, the frequencies of the two modes will split.
Zambelli and Stringari used the sum rule approach to obtain an
analytical result of this splitting when a single vortex lies in
the center of the condensate\cite{Stringari}. Svidzinsky and
Fetter also obtained a consistent result by solving the the
hydrodynamic equations with perturbational
approximation\cite{Fetter}. They both found that the frequency
splitting is proportional to the total angular momentum carried by
the condensate. Later the frequency splitting when the vortex is
off-center was also calculated
pertubatively\cite{Fetter-review}\cite{offaxisvortex}. They found
that the result obtained from the sum rule approach does not agree
with that from the perturbative approach any more. In this
section, we will first calculate the collective mode splitting in
presence of a vortex dipole from the perturbation approximation,
and we find the similar disagreement between perturbation approach
and the sum rule approach will also occur here. We will briefly
discuss the reason why the sum rule approach fails to meet the
result of perturbative approach. We should point out that it is
not the sum rule itself but the incorrect way we use it that leads
to this failure.

\subsection{Frequency splitting induced by vortex dipole}
Substituting $\psi=\sqrt{\rho}e^{i\phi}$ into Eq.(\ref{G-P}), we
could obtain the well known hydrodynamic equations in the T-F
region.
\begin{eqnarray}
\frac{\partial\rho}{\partial{t}}+\vec{\nabla}\cdot(\rho\vec{\upsilon})=0,\\
M\frac{\partial\vec{\upsilon}}{\partial{t}}+\vec{\nabla}(V+g\rho-\mu+\frac{1}{2}M\upsilon^2)=0,
\end{eqnarray}
where $\vec{\upsilon}=\vec{\nabla}\phi$.  Expanding
$\rho=\rho_0+\delta\rho$ and
$\vec{\upsilon}=\vec{\upsilon}_0+\delta\vec{\upsilon}$ with
\begin{equation}
\vec{\upsilon}_0=\vec{\upsilon}_1+\vec{\upsilon}_2
=\frac{\hat{z}\times(\vec{r}-\vec{r}_1)}{\left|\vec{r}-\vec{r}_1\right|^2}
-\frac{\hat{z}\times(\vec{r}-\vec{r}_2)}{\left|\vec{r}-\vec{r}_2\right|^2},
\end{equation}
and after linearization, we could obtain the equation for
collective modes
\begin{equation}
(\omega^2+\frac{g}{M}\vec{\nabla}\cdot\rho_0\vec{\nabla})\delta\rho
=-i\omega\vec{\nabla}(\vec{\upsilon}_0\delta\rho)+\frac{ig}{M\omega}\vec{\nabla}
\cdot[\rho_0\vec{\nabla}(\vec{\upsilon}_0\cdot\vec{\nabla})\delta\rho].\label{linearization}
\end{equation}
We notice that the left hand of Eq.(\ref{linearization}) is just
the equation of collective motion corresponding to a vortex-free
ground state. And we also note that the effect of $\upsilon_0$ to
the collective modes is always small, because $\upsilon_0$
decreases quickly from the vortex dipole center as ${1}/{r^2}$
while $\delta\rho$ increases from the condensation center as
${r^m}$, their overlapping is always small except the position of
the vortex dipole is nearby the boundary. So we could treat the
right hand side as perturbation safely and obtain
\begin{equation}
\omega^2-\omega_0^2=-i\omega_0\langle g|\vec{\upsilon}_0
\cdot\vec{\nabla}|g\rangle+\frac{ig}{M\omega_0}\langle
g|\vec{\nabla}\cdot\rho_0\vec
{\nabla}(\vec{\upsilon}_0\cdot\vec{\nabla})|g\rangle,\label{splitting}
\end{equation}
where $\omega_{0}$ is the collective mode frequency in a vortex
free case. Hereafter we consider the following surface mode
\begin{equation}
|g\rangle=\sqrt{\frac{m+1}{\pi}}\frac{r^m}{R^{m+1}}e^{i\theta}.
\end{equation}
From the calculation presented in Appendix C, we could obtain the
splitting between $+m$ mode and $-m$ mode as
\begin{equation}
\omega_+-\omega_-=\frac{2(m+1)}{R^2}\left[\left(\frac{r_2}{R}\right)^{2m-2}
-\left(\frac{r_1}{R}\right)^{2m-2}\right].\label{result}
\end{equation}
In experiments, from the optical absorption image the vortices
positions $r_{1}$ and $r_{2}$ can be determined, and then by
measuring the frequency splitting and comparing the obtained
result with Eq.(\ref{result}), one can distinguish a vortex dipole
from two vortices with the same winding directions.

\subsection{Discussion on the Application of Sum Rule}

When we apply the formulism of Ref.\cite{Stringari} to calculate
the frequency splitting of $\pm m$ modes, we can not obtain the
same result as Eq.(\ref{result}). To understand the reasons, let
us first briefly remind ourself of the sum rules approach. The
general formula of the sum rules for an arbitrary operator
$\mathcal{F}$ can be easily proved as
\begin{eqnarray}
\sum_n(E_n-E_k)\left|\mathcal{F}_{nk}\right|^2=\frac{\langle k|[\mathcal{F^+},[H,\mathcal{F}]]|k\rangle}{2},\\
\sum_n(E_n-E_k)^3\left|\mathcal{F}_{nk}\right|^2=\frac{\langle
k|[[\mathcal{F^+},H],[H,[H,\mathcal{F}]]]|k\rangle}{2}.
\end{eqnarray}
If we choose $k=0$, and with a particular choice of the operator
$\mathcal{F}$ which satisfies
$\left|\mathcal{F}_{n0}\right|\sim{\delta_{n0}}$, we could obtain
\begin{equation}
\hbar^2\omega_n^2=\frac{\langle
0|[[\mathcal{F^+},H],[H,[H,\mathcal{F}]]]|0\rangle}{\langle
0|[\mathcal{F^+},[H,\mathcal{F}]]|0\rangle}\label{sumrule}.
\end{equation}
This is a powerful method because we can obtain the excitation
energy directly from the property of the ground state without the
explicit wavefunction of excitation state. However, the successful
application of the sum rule approach relies on the proper choice
of $\mathcal{F}$, which should excite the ground state to only one
definitive excited state. When the system possesses some
particular symmetry, we may easily find out a suitable
$\mathcal{F}$. That's true when there is a central vortex, as
rotation invariance is still present and the excited state should
also have the form $e^{im\phi}$. So we could easily select
$\mathcal{F}$ as $(x\pm{iy})^2$ to make use of the sum rule.
However, the rotation invariance is broken in the case of vortex
dipole. The choice of $\mathcal{F}$ still as $(x\pm{iy})^2$ will
excite several states, and Eq.(\ref{sumrule}) will not tenable.
That's why the sum rule approach fails to agree with the
perturbative result in Ref.\cite{offaxisvortex}.

\section{Conclusion}
In summary, by applying a convenient integration method in complex
coordinate, we have obtained the fully analytic expressions of the
angular momentum and energy of a vortex dipole in a trapped BEC.
We also suggest an effective method to generate and detect the
vortex dipole. Our work may provide a starting point for the
research in the problem of vortex dipole excitation. There still
remains many interesting fundamental issues in both theoretical
and experimental physics, such as its dynamics, stability and
detailed generation mechanism.

The precession of off-center vortex in a trapped condensate has
been fully investigated both in experiment and theory. The mutual
interaction between vortex and antivortex is presumed to bring a
more nontrivial trajectory for vortex dipole. On the other hand,
its dynamic behavior is closely related to its stability. In the
dual electromagnetic picture\cite{Aop}, the positive and negative
charge, to which the vortex dipole corresponds, will tend to
annihilate each other via electromagnetic radiation, say phonon
excitation in original picture. However, in a trapped condensate,
there are two mechanism to stabilize the vortex dipole, one is the
angular momentum conservation, and the other is the discrete
spectrum of the the phonon in a finite system. Investigation of
this interesting competition will reveal the underlying physics of
the numerical simulation\cite{stablevortexdipole}. The fundamental
mechanism of generating vortex dipole from a local stirring is
even more complex, and it may resemble the birth of electron and
positron from the vacuum polarization of the electromagnetic
field.

Another important issue which should also arouse great attention
is the complex phase diagram of rotating two-dimensional
superfluid. When the angular momentum L equals zero, the critical
temperature $T_c$ for K-T transition is well understood. But when
the superfluid carries a certain angular momentum, the dependence
of $T_c$ to the angular momentum, as far as we are concerned,
hasn't been clearly revealed. On the other hand, with the increase
of angular momentum, the vortex dipoles will break into free
vortices, and finally stable vortex excitation will become
dominative. However, the explicit mechanism of the transition from
vortex dipole excitation to stable vortex excitation is also still
unknown. The interplay of the angular momentum and the thermal
excitation will bring rich physics to the rotating two-dimensional
superfluid. Several works are proceeding.

\textit{Acknowledgements}: The authors should like to thank
Professor C. N. Yang for encouragement. And the authors would like
to acknowledge for helpful discussions with Z.Y. Weng, L. Chang,
R. L$\ddot{u}$ and X.L.Qi. QZ would like to thank Professor T.L.Ho
and Professor Z.Y.Weng for their valuable support during his
predicament excruciated by visa delay. QZ also thanks CASTU for
providing a friendly environment to finish this work. This work is
supported by National Natural Science Foundation of China ( Grant
No. 10247002 )

\appendix
\section{Integration in Complex Coordinate and Green Formula}
In this section we introduce our method used to do integration in
a two-dimensional space. By denoting $z=x+iy$ and $\bar{z}=x-iy$,
we can turn the integration into the complex coordinate as
\begin{equation}
\int_\Omega{f(x,y)}dxdy=-\frac{1}{2i}\int_\Omega{f(z,\bar{z})}dz\wedge{d\bar{z}},
\end{equation}
where $\Omega$ denotes the integration region, and the $\wedge$
operate fulfils
$dx_i\wedge{dx_j}=(\delta_{ij}-1)dx_j\wedge{dx_i}$, here $x_i$ is
the coordinate in real or complex space. Introducing a
differential $1$-form $\omega=g(z,\bar{z})dz$ in which the
function $g$ satisfies
\begin{equation}
\frac{\partial{g(z,\bar{z})}}{\partial{z}}=f(z,\bar{z}),
\end{equation}
we have $d\omega=f(z,\bar{z})dz\wedge{d\bar{z}}$. According to the
Green formula, we could change the above integration over $\Omega$
to over its boundary denoted as $\partial\Omega$ as
\begin{equation}
\int_\Omega{f(z,\bar{z})}dz\wedge{d\bar{z}}=\int_{\partial\Omega}g(z,\bar{z})d\bar{z}
\end{equation}
This method is very useful when doing integration in a irregular
region, because it turns a two-dimensional integration into a
one-dimensional one, and result of the latter is easier to be
obtained numerically or approximatively.

\section{Vortex Dipole Energy}
\begin{figure}[htbp]
\begin{center}
\includegraphics[width=1.2in]
{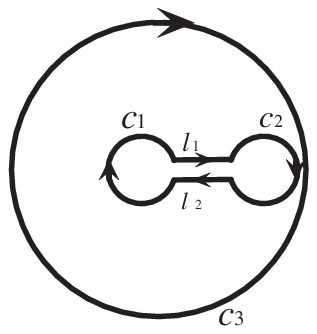}
\includegraphics[width=1.2in]
{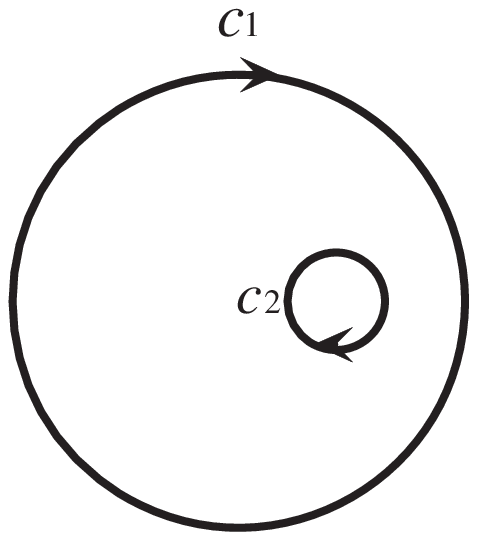} \caption{The integration along the boundary in the complex
coordinates.} \label{fig:integration1}
\end{center}
\end{figure}

To obtain the result of the integration Eq.(\ref{Dipoleenergy}) we
first calculate
\begin{equation}
\int_\Omega\frac{\left|z_1-z_2\right|^2}{\left|z-z_1\right|^2\left|z-z_2\right|^2}d\bar{z}\wedge{dz}.
\end{equation}
According to above method, it turns to
\begin{equation}
\int_{\partial\Omega}\frac{z_1-z_2}{(z-z_1)(z-z_2)}\ln\frac{\bar{z}-z_1}{\bar{z}-z_2}dz\label{B1}
\end{equation}
We notice that not only does the integrand has two singularities
$z_1$ and $z_2$, but also includes multi-value parts
$\ln(\bar{z}-z_1)$ and $\ln(\bar{z}-z_2)$, so the integration in
the boundary should
be as the left side of Fig.(\ref{fig:integration1}).\\

In the condition that $d\gg \Lambda$, the integration along circle
$C_1$ reads
\begin{eqnarray*}
&&\int_{z-z_1=\Lambda{e^{i\varphi}}}\frac{z_1-z_2}{(z-z_1)(z-z_2)}\ln\frac{\bar{z}-z_1}{\bar{z}-z_2}dz\\
&&\simeq\int_0^{2\pi}\frac{-d}{(z-z_1)(-d)}[\ln\frac{\Lambda}{d}-i(\varphi+\pi)]i(z-z_1)d\varphi\\
&&=-2i\pi\ln\frac{\Lambda}{d}
\end{eqnarray*}
The similar result of integration along $C_2$ circle reads
$2i\pi\ln\frac{d}{\Lambda}$. We notice that the imaginary part of
the logarithmic function in integrand in branch $l_{1}$ differs
from that in $l_{2}$ by $2\pi$, the integration along the two
branches gives
\begin{equation}
\int_\Lambda^{d-\Lambda}\frac{-d}{x(x-d)}2\pi{dx}=4i\pi\ln\frac{d}{\Lambda}.
\end{equation}
The integration is more complex along circle $C_3$, we shall
decompose Eq.(\ref{B1}) as
\begin{equation}
\sum_{i=1,2}\int\frac{\ln(z-r_i)}{\bar{z}-r_i}d\bar{z}-\sum_{i\neq{j}}\int\frac{\ln(z-r_j)}{\bar{z}-r_i}d\bar{z}.
\end{equation}
The general form of the integrand could be written as
\begin{equation}
\frac{1}{\bar{z}}\left[1+\sum_{m=1}^{\infty}(\frac{z_j}{\bar{z}})^m\right]
\left[\ln{z}-\sum_{n=1}^{\infty}(\frac{z_i}{\bar{z}})^n\right],
\end{equation}
then each of the four terms after decomposition is easier to be
calculated. We obtain the integration along $C_3$ circle as
\begin{equation}
2i\pi\left[\ln\frac{\left(1-\frac{\left|z_1\right|^2}{R^2}\right)\left(1-\frac{\left|z_2\right|^2}{R^2}\right)}
{\left(1-\frac{\left|z_1z_2\right|}{R^2}\right)^2}\right].
\end{equation}
The above value will vanish in the limit of $d/R\rightarrow0$,
thus could be neglected. Then we shall obtain
\begin{equation}
\int_\Omega\frac{\left|z_1-z_2\right|^2}{\left|z-z_1\right|^2\left|z-z_2\right|^2}d\bar{z}\wedge{dz}
=8i\pi\ln\frac{d}{\Lambda}.
\end{equation}
Proceeding the similar steps, we could obtain
\begin{equation}
\int_\Omega\frac{\left|z\right|^2\left|z_1-z_2\right|^2}{R^2\left|z-z_1\right|^2\left|z-z_2\right|^2}d\bar{z}\wedge{dz}
=8i\pi\frac{D^2}{R^2}\ln\frac{d}{\Lambda}.
\end{equation}
The energy of a vortex dipole will finally read
\begin{equation}
E=\frac{2\hbar^2\rho_0}{M}\left[1-\left(\frac{D}{R}\right)^2\right]\ln\frac{d}{\Lambda}.
\end{equation}

\section{Frequency Splitting}

We notice that the Eq.(\ref{splitting}) is linear with respect to
$\upsilon_0$, we could first calculate a single positive vortex
case. That is
\begin{equation}
\upsilon_x=-\frac{z'-\bar{z}'}{2iz'\bar{z}'},
\end{equation}
and
\begin{equation}
\upsilon_y=\frac{z'+\bar{z}'}{2z'\bar{z}'},
\end{equation}
where $z'=z-z_1$, $z_1$ denotes the location of vortex. Thus
\begin{eqnarray}
&&\langle g|\vec{\upsilon}_0\cdot\vec{\nabla}|g\rangle =\langle
g|\frac{\partial_z}{\bar{z}-zz_1}-\frac{\partial_{\bar{z}}}{z-z_1}
|g \rangle \nonumber\\
&&=-\frac{m+1}{\pi{R^{2m+2}}}\int_\Omega\frac{\bar{z}^m}{\bar{z}-z_1}\partial_zz^mdz\wedge{d\bar{z}}\nonumber\\
&&=-\frac{m+1}{\pi{R^{2m+2}}}\int_{\partial\Omega}\frac{\bar{z}^mz^m}{\bar{z}-z_1}d\bar{z}
\end{eqnarray}
The integration along the boundary is expressed as the right ride
of Fig(\ref{fig:integration1}), and the result is easily to be
obtained as
\begin{equation}
\langle
g|\vec{\upsilon}_0\cdot\vec{\nabla}|g\rangle=\frac{(m+1)i}{R^2}\left(1-\frac{|z_1|^{2m}}{R^{2m}}\right).
\end{equation}
With the same process, we could also obtain $\langle
g|\vec{\nabla}\cdot\rho_0\vec{\nabla}(\vec{\upsilon}_0\cdot\vec{\nabla})|g\rangle$
as
\begin{equation}
-\frac{2im(m+1)\rho_0}{R^4}\left[1+\left(\frac{|z_1|}{R}\right)^{2m}-2\left(\frac{|z_1|}{R}\right)^{2(m-1)}\right].
\end{equation}
Recalling that $\omega_0=\omega_\bot\sqrt{m}$ and substituting the
above expression back into Eq.(\ref{splitting}), we will obtain
\begin{equation}
\omega-\omega_0=\frac{m+1}{R^2}\left[1-\left(\frac{|z_1|}{R}\right)^{2(m-1)}\right].
\end{equation}
The same result of Ref.(\cite{offaxisvortex}) is again obtained.
As we have emphasized that Eq.(\ref{splitting}) is linear with
$\upsilon_0$, the above result of single vortex can be directly
extended to vortex dipole, that is
\begin{equation}
\omega-\omega_0=\frac{m+1}{R^2}\left[
\left(\frac{|z_2|}{R}\right)^{2(m-1)}-\left(\frac{|z_1|}{R}\right)^{2(m-1)}\right],
\end{equation}
where $z_2$ is the position of anti-vortex. To calculate the
frequency of $-m$ mode, we only need to change the sign of
$\upsilon_0$ equivalently. So the frequency splitting of $\pm{m}$
mode is
\begin{equation}
\omega_+-\omega_-=\frac{2(m+1)}{R^2}\left[\left(\frac{|z_2|}
{R}\right)^{2(m-1)}-\left(\frac{|z_1|}{R}\right)^{2(m-1)}\right].
\end{equation}

\end{document}